# Piezo-Voltage Manipulation of the Magnetization and Magnetic Reversal in Thin Fe Film

Y. Y. Li, W. G. Luo, L. J. Zhu, J. H. Zhao, and K. Y. Wang[1]

*SKLSM, Institute of Semiconductors, CAS, P. O. Box 912, 100083 Beijing, People's Republic of China*

**Abstract**

We carefully investigated the in-plane magnetic reversal and corresponding magnetic domain structures in Fe/GaAs/piezo-transducer heterostructure using longitudinal magneto-optical Kerr microscopy. The coexistence of the <100> cubic magnetic anisotropy and [$1\bar{1}0$] uniaxial magnetic anisotropy was observed in our Fe thin film grown on GaAs. The induced deformation along [110] orientation can effectively manipulate the magnetic reversal with magnetic field applied along magnetic uniaxial hard [110] axes. The control of two-jump magnetization switching to one-jump magnetization switching during the magnetic reversal was achieved by piezo-voltages with magnetic field applied in [100] direction. The additional uniaxial anisotropy induced by piezo-voltages at ±75 V are ±1.4×10$^3$ J/m$^3$.

---

[1] Corresponding author's E-mail: kywang@semi.ac.cn

# I. INTRODUCTION

Electric field control of the ferromagnetic magnetization process has become increasingly important because it has great potential applications for the memory device, magnetic logic, magneto-electric sensor and the integration of magnetic functionalities into electronic circuits [1,2]. The approaches adopted to achieve this aim including the electric field gating to change the magnetic anisotropy of an ferromagnetic semiconductors or ultrathin ferromagnetic metals [3,4], electric current generating spin-orbit coupling field to assist the magnetization reversal [5-7], ultrafast polarized optics to switch the magnetization [8], and strain to change the anisotropy and the magnetization reversal of magnetic thin film [9,10]. Since the magneto crystalline anisotropy is originated from the spin-orbital coupling of the materials, modification of lattice constant using piezo-voltage can thus directly manipulate the magnetic anisotropy and the magnetization reversal. For decades, the interest of Fe films anisotropy research never diminishes because of its broad promising application [11]. The thin Fe film grown on GaAs substrate has the Curie temperature well above the room temperature and it also integrates ferromagnetism and traditional semiconductor GaAs together [12,13]. Very rich magnetic anisotropy, coexistence of the in-pane cubic and uniaxial anisotropy, was observed in the epitaxial growth of thin Fe film on GaAs substrate [14,15]. Thus understanding the physics of piezo-voltage control of the magnetic anisotropy and magnetic reversal of this system could be very important for realization the future metallic-semiconducting spintronic applications [16,17].

In this paper, we firstly investigated the magnetic hysteresis and the corresponding magnetic domain structures during the magnetic reversal in the virgin state of the Fe/GaAs/piezo-transducer heterostructure. We then compared the hysteresis loops of the Fe thin film with magnetic field applied in-plane different crystalline orientations under positive/negative piezo-voltages. The piezo-voltages can effectively control the magnetization crossing the intermediate magnetic state during the magnetic reversal. Also the magnitude of the additional uniaxial anisotropy induced by piezo-voltages was obtained to be $\pm 1.4 \times 10^3$ J/m$^3$ at $\pm 75$ V.

# II. EXPERIMENTS

The ultra-thin 5 nm Fe was grown on n-type GaAs/GaAs substrate using molecular-beam epitaxiy (MBE). The substrate of the sample was first polished to 120±10 μm to ensure that the deformation can be transferred to the sample. Then the sample was bonded to the lead zirconate titanate (PZT) piezotransducer using two-component epoxy after thinning the substrate. The positive/negative voltage produces a uniaxial tensile/compressive strain in the direction of [110]. The induced strain/stress along [110] orientation was measured by strain gauge, which was found to be linearly changed with the applied voltage. The magnitude of the additional uniaxial strain/stress for a piezo-voltage of +/-75 V is approximately +/-3.5×10$^{-4}$. Under zero and applied piezo-voltages, the magnetization vectors and the corresponding magnetization domains of the thin Fe film during magnetization reversal along in-plane different orientations were measured by using longitudinal magneto-optical Kerr microscopy (LMOKM). In the longitudinal setup, the Kerr rotation angle is proportional to the magnitude of the magnetization component along the projection direction of the incident light in the plane. The magnetization vectors thus can be used to determine the relative magnitude ($M/M_S$) (where $M$ is the magnetization component along the magnetic field direction and $M_S$ is the saturation magnetization) and direction of the magnetization. Thus, the hysteresis loops with magnetic field along different in-plane orientations measured by the longitudinal Kerr effect can be used to investigate the in-plane magnetic anisotropy. In this work, the experimental measurements were all performed at room temperature, and the magnetic field is at a fixed frequency of 0.5 Hz.

### III. RESULTS AND DISCUSSIONS

#### A. Magnetic anisotropy in virgin state

Rotating the sample in the plane, the LMOKM was used to measure the hysteresis loops and the corresponding magnetic domain structures with magnetic field applied in any direction in plane. Figs.1(a)-(c) show the Kerr rotation angle during the magnetization reversal with field applied close to the in-plane major crystalline [110], [1$\bar{1}$0] and [100] orientations, respectively. Each curve was measured deviating 5$^0$ from the main crystalline to make sure the magnetization procedure route towards counterclockwise.

As shown in Figs. 1(a) and (b), although only one step magnetization switching was observed during magnetization reversal with magnetic field applied in both [110] and [1$\bar{1}$0] directions. However, the inside physics of magnetic reversal in these two directions are very different. The remnant Kerr rotation angle is close to the saturation Kerr rotation angle (about 40 mdeg) with the magnetic field applied in [1$\bar{1}$0] direction. Also the very square and sharp magnetic hysteresis loop was observed along the $[1\bar{1}0]$ orientation, indicating the easiest magnetic anisotropy orientation. However, the remnant Kerr rotation angle in [110] direction is only 20 mdeg and the magnetization reversal is much less abrupt, indicating the relative hard magnetic anisotropy orientation. With magnetic field sweeping from positive to negative in $[1\bar{1}0]$ direction, the magnetization directly switches from the initial $[1\bar{1}0]$ orientation single domain state crossing the 180 degree energy barrier to the final $[\bar{1}10]$ magnetic state, which is confirmed by the corresponding magnetic domain structures in Fig. 1(b). The system is only broken into magnetic domains mentioned as II and IV during magnetization switching in Fig. 1(b). However, the magnetization along [110] orientation is different from that of the $[1\bar{1}0]$ orientation during the magnetization reversal. Sweeping the magnetic field (the amplitude is larger than saturated magnetization field) from positive to negative in [110] direction, the magnetization firstly coherently rotates from the field direction to the easy axis direction, then the system breaks into domains and prorogates through $180^0$ domain walls at $H_C$, after that the magnetization coherently rotates again from the easy axis to the external magnetic field direction. The strong asymmetry of magnetic hysteresis between [110] and $[1\bar{1}0]$ orientation suggests that there is a uniaxial magnetic anisotropy between these two directions, which originates from the surface reconstruction of Fe grown on GaAs [13,14].

As shown in Fig. 1(c), a two-step-jump magnetic switching behavior was observed during magnetic reversal with magnetic field applied along [100] orientation. Very similar behavior was also observed in [010] orientation. This symmetry is attributed to the four fold in-plane magnetic anisotropy. With magnetic field sweeping from positive to negative in [100] direction, the magnetization firstly switched anticlockwise to the intermediate state rather than directly switched from the initial state crossing a 180 degree barrier to the final state. The magnitude of the Kerr rotation angle of the intermediate state is about $\sqrt{2}/2$ to that of the initial state for the [100] direction hysteresis loop, which suggests that the intermediate state is along $[1\bar{1}0]$ direction and about $45^0$ away from the initial state. So when the external field along the [100] orientation sweeping from positive to negative through zero field, the free energy minima was along [100], $[1\bar{1}0]$, $[\bar{1}00]$ orientation, respectively. The corresponding domain states during the magnetization reversal were shown in the inserted images of

Fig. 1(c). The scanning area of the magnetic domain images is 580 μm×580 μm. The three plateaus (marked as I, III and V) from positive to negative represent the single domain state of magnetization along [100], [1$\bar{1}$0] and [$\bar{1}$00], respectively. While the magnetization of film breaks into domains in the two sharp transitions (II and IV) in Fig. 1(c), which is corresponding to the low and high two-step switching coercive fields $H_{C1}$ and $H_{C2}$, respectively.

The coexistence of the <100> cubic magnetic anisotropy and [1$\bar{1}$0] uniaxial magnetic anisotropy was found in our Fe thin film grown on GaAs substrate [13]. Similar magnetic anisotropy was also observed in Fe grown on InAs substrate and GaMnAs grown on GaAs substrate [18,19]. Considering the <100> cubic magnetic anisotropy and [1$\bar{1}$0] uniaxial magnetic anisotropy, the magnetic free energy of our system under external magnetic field can be written down as:

$$E = \frac{1}{8}H_1 M_S cos^2 2\theta + \frac{1}{2}H_2 M_S cos 2\theta - M_S H cos(\theta - \varphi) \qquad (1)$$

where $H_1$ is the cubic anisotropy anisotropic field, $H_2$ is the uniaxial anisotropy field, $M_S$ is the saturated magnetization of Fe film which is about to 1700 A/m [20], $\theta$ is the angle between the magnetization and hard axis [110] orientation and the $\varphi$ is the angle between the applied field and [110] direction. According to the Stoner-Wohlfarth formula, at a given applied field, the magnetization coherently follows the global free energy minimum: $\partial E / \partial \theta = 0$. This formula can well describe the magnetization reversal in [110] direction with the magnitude of the magnetic field above 20 Oe, as shown in the insert of Fig. 2(a). With $cos\theta = M/M_S$, the relationship between the anisotropic fields (or anisotropy) and external magnetic field H applied in the [110] direction can be written down as:

$$H = 2H_1(M/M_S)^3 + (H_2 - H_1)(M/M_S) \qquad (2).$$

By fitting the coherent rotation part of the magnetization reversal with magnetic field applied in [110] orientation, the anisotropic fields thus can be calculated. The cubic anisotropic and uniaxial anisotropic fields are 356 Oe and 100 Oe in the virgin state. However, this formula can't be used to describe the magnetization reversal with magnetic field applied in both [110] and [1$\bar{1}$0] orientations. The magnetization reversal was confirmed to not rotate coherently with sweeping the external magnetic field along [110] and [100] orientations, proved by the domain images in Figs. 1(b) and (c). The magnetization switching will happen if the gaining energy of magnetic domains is larger than the energy barrier of the two neighbor local minimas. Cowburn et al. has developed a magnetization reversal model considering both the $90^0$ and $180^0$ domain walls [21]. From the obtained magnetic anisotropy above by fitting the coherent rotation in [110] orientation, the two global energy minima directions of this sample are [1$\bar{1}$0] and [$\bar{1}$10] directions. Considering the energy barrier between these two global energy minimas, at coercive field this energy barrier of the $180^0$ domain wall is equal to the pining field $\varepsilon_{180^0}$. The pinning energy can be written down as: $\varepsilon_{180^0} = 2M_S H_{C,C2} \sin\varphi$, where $H_C$ is the coercivity when the external field near the uniaxial easy axis. In order to obtain the pinning field, the angular dependence of the coercive field was measured, which was plotted in Fig.2 (b). By fitting the angular dependence of coercive field $H_{C2}$, the energy barrier $\varepsilon_{180^0}$ = 560 J/cm$^3$ through the hard axis [110]/[$\bar{1}\bar{1}$0] was obtained.

### B. Piezo-voltage induced magnetic anisotropy

With the piezo-voltage induced deformation along [110] direction, we carefully investigated the in-plane magnetic hysteresis under strained or stressed situation of the Fe/GaAs thin film. Figs. 3(a)-(c) show the Kerr rotation angle during the magnetization reversal at strain/stress (+/-75 V) states with field applied in [110], [1$\bar{1}$0] and [100] orientations with $5^0$ deviation, respectively. For reference, the magnetic hysteresis without strain is shown in Fig. 3 as well. It is very interesting that we can efficiently manipulate the magnetization process with piezo-voltages. With magnetic field applied in [110] orientation, the magnetic hysteresis loops shows very clear difference close to $H_C$, which is shown in Fig. 3(a). Compared with U = 0 V, the magnetic hysteresis becomes more square and the coercive field increases with U = -75 V. However, the magnetic hysteresis becomes less square and the coercivity decreases with U = 75 V. This is due to piezo-voltages induced additional uniaxial anisotropy, which widens the 1-jump region about the uniaxial easy direction and diminishes the l-jump region about the uniaxial hard direction [20]. However, the magnetic hysteresis loops show no clear difference with magnetic field along [1$\bar{1}$0] orientation. This could be the piezo-voltages induced energy which is insignificant to the magnitude of the uniaxial anisotropy $K_2$. As shown in Fig. 3(c), the magnitude of Kerr rotation angle in the three plateaus was unvaried under piezo-voltages with magnetic field applied in [100] direction, while the most pronounced changes of the two sharp magnetization switches was observed with applied piezo-voltages. The coercive fields of the two sharp changes $H_{C1}$ and $H_{C2}$ correspond to the magnetization crossing from [100] to [1$\bar{1}$0] ([$\bar{1}$00] to [$\bar{1}$10]) and from [1$\bar{1}$0] to [$\bar{1}$00] ([$\bar{1}$10] to [100]), respectively. The $H_{C2}$ increases with U = 75 V, while it decreases with U = -75 V. With sweeping magnetic field from positive to negative in [100] direction, the first switching of $H_{C1}$ was happened much early at positive field with U = 75 V. However, the $H_{C1}$ increases dramatically and the first step switching nearly matches with the second step switching with U= -75 V. The relative change of the $H_{C1}$ and $H_{C2}$ depends not only on the exact direction of applied magnetic field and also the magnitude of the strain/stress. Figs. 3(a) and (c) show that the most effectively manipulation by the strain and stress is the magnetic switching from [1$\bar{1}$0] to [$\bar{1}$10] ([$\bar{1}$10] to [1$\bar{1}$0]) direction with field applied in [110] direction and the 180 degree switching from [100] to [1$\bar{1}$0] ([$\bar{1}$00] to [$\bar{1}$10]) direction with field applied in [100] orientation. This is attributed to contribution of an extra uniaxial anisotropy induced by the positive/negative piezo-voltages induced the strain/stress, which will enhance/decrease [1$\bar{1}$0] uniaxial magnetic anisotropy, corresponding to increase/decrease the energy barrier along [110] orientation.

With magnetic field applied in [110] orientation, the magnetic hysteresis loops were measured with applied different piezo-voltages, which is shown in Fig. 4(a). The magnetic hysteresis becomes more and squarer and the coercive field increases with increasing the magnitude of the negative piezo-voltages (stress). However, the magnetic hysteresis becomes less and less abrupt and the coercivity decreases with increasing the strain. With piezo-voltage increased from -75 to 75 V (stress to strain), the corresponding coercive field decreases from 1.6 to 0.8 Oe. Compared the coercivity at U = +/-75 V to the virgin state, the strain/stress induced the change of the coercive fields along this orientation is up to 30%. The coercivity was found to roughly decrease linearly with increasing the piezo-voltages from -75 to 75 V, which is shown in Fig. 4(b). This phenomenon is attributed to the piezo-voltages induced the extra uniaxial anisotropy.

When there is a deformation along [110] orientation, the induced additional uniaxial anisotropic term will be superimposed on the magnetic free energy. In this work the additional uniaxial term induced by strain/stress has the same symmetry as $K_2$. Then the magnetic free energy of Eq. (1) can be modified as: $E = \frac{1}{8}H_1 M_S \cos^2 2\theta + \frac{1}{2}H_2 M_S \cos^2 \theta + \frac{1}{2}H_a M_S \cos^2 \theta - M_S H \cos(\theta - \varphi)$ (3)

where $H_a$ is the additional uniaxial anisotropic field induced by the deformation along [110]

orientation. Thus the relationship between the anisotropic fields and external magnetic field H applied in the [110] direction with the extra uniaxial anisotropic field can be written down as:

$$H = 2H_1(M/M_S)^3 + (H_2 + H_a - H_1)(M/M_S) \qquad (4).$$

Using the virgin state fitted cubic and uniaxial anisotropy $H_1$ = 356 and $H_2$ = 100 Oe, then the piezo-voltages induced uniaxial anisotropy can be obtained by fitting the coherent rotation magnetization with magnetic field applied in [110] direction using Eq. (4), which is shown in Fig. 5. The additional uniaxial anisotropic fields under strain/stress are +/- 8 Oe with the piezo-voltage +/-75 V respectively, which is about 8% of the magnitude of $H_2$. Thus the additional uniaxial anisotropy induced by piezo-voltages at +/- 75 V are +/-1.4×$10^3$ J/$m^3$.

## IV. CONCLUSION

In summary, the magnetic anisotropy and magnetic reversal of Fe/GaAs/PZT hetero-structure with and without piezo-voltages were carefully investigated using longitudinal magneto-optical Kerr microscopy. The coexistence of cubic anisotropy and in-plane uniaxial anisotropy in [110] direction was found in the virgin state. The system was found to break into domains only at sharp magnetization switching regime, which was confirmed by the corresponding magnetic domain structures. With piezo-voltages induced deformation in [110] orientation, the strain/stress was found to effectively manipulate the magnetization reversal. The coercivity during the magnetic reversal was found to roughly decrease linearly with increasing the piezo-voltages from -75 to 75 V with magnetic field applied in [110] direction. The two jump magnetization switching to one jump magnetization switching during the magnetic reversal was achieved by piezo-voltages with magnetic field applied in [100] direction. The additional uniaxial anisotropy induced by piezo-voltages at +/- 75V are +/-1.4×$10^3$ J/$m^3$, which is large enough to control the magnetization reversal in our Fe/GaAs/PZT hetero-structure.

## ACKNOWLEDGEMENTS

We thank X. Q. Ma for useful discussions. We also thank Shanghai Synchrotron Radiation Facility for characterizing FCC structure of the Fe film. This work was supported by "973 Program" No. 2011CB922200, NSFC Grant Nos. 11174272 and 61225021. K.Y.W. acknowledges the support of Chinese Academy of Sciences "100 talent program".

**Figure Captions:**

*FIG.1.* Magnetic hysteresis loops in the virgin state with magnetic field applied along in-plane (a) [110] orientation, (b) [1$\bar{1}$0] orientation, and (c) [100] orientation. The insert of (a) shows the magnetic hysteresis loop in a wide field range, the insert images of (b) and (c) corresponding to the magnetic domain structures during the magnetic reversal in [1$\bar{1}$0] and [100] orientations, respectively. The scanning area of the magnetic domain images is 580 μm×580 μm.

*FIG.2.* (a) The experimental and the fitting of coherent magnetization with magnetic field in [110] orientation; (b) Angular dependent of the coercive fields with magnetic field applied along different in-plane orientations in the virgin state, where θ=0º, along [110] direction. The areas marked I is the areas with one-step hysteresis loop, and the areas marked II is the areas with two-step hysteresis loop.

*FIG.3.* Magnetic hysteresis loops under piezo-voltages at 0 V (black curve), 75 V (red curve) and -75 V (blue curve) with magnetic field applied along (a) [110] direction, (b) [1$\bar{1}$0] direction, and (c) [100] direction, respectively.

*FIG.4.* (a) The magnetic hysteresis loops with sweeping magnetic field along [110] direction at different piezo-voltages; (b) The coercivity dependence of the piezo-voltages with magnetic field applied in [110] direction, the line is guide to the eye.

*FIG.5.* The experimental data with field applied in [110] direction in the coherent rotation with piezo-voltages at 75 V (open square) and -75 V (open triangle), respectively. The lines are fitting curves.

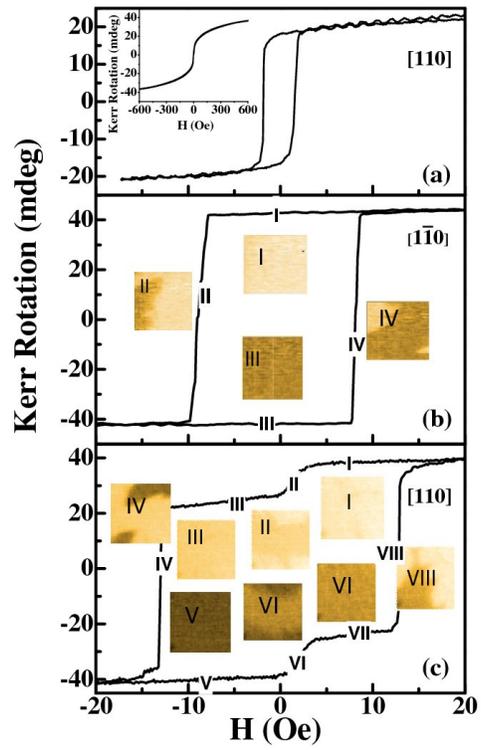

*Figure 1 Y.Y. Li et al.*

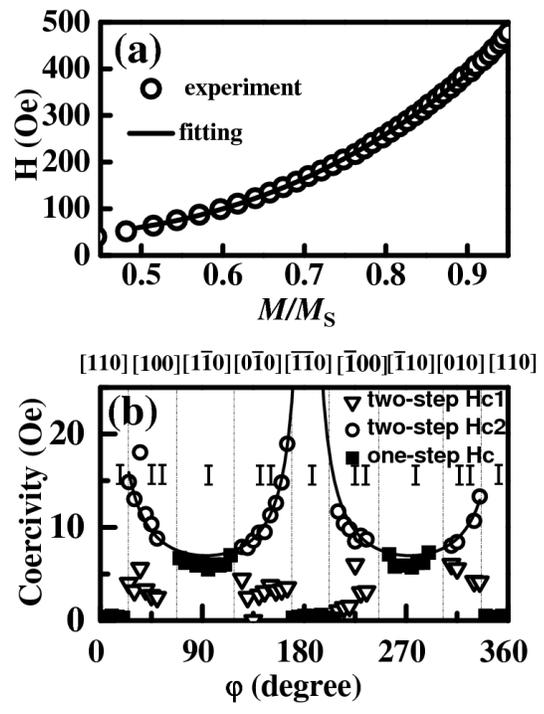

*Figure 2 Y.Y. Li et al.*

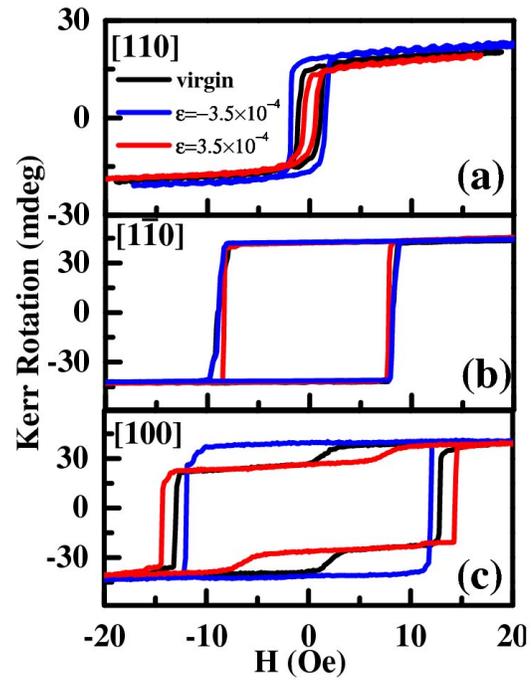

Figure 3 Y.Y. Li et al.

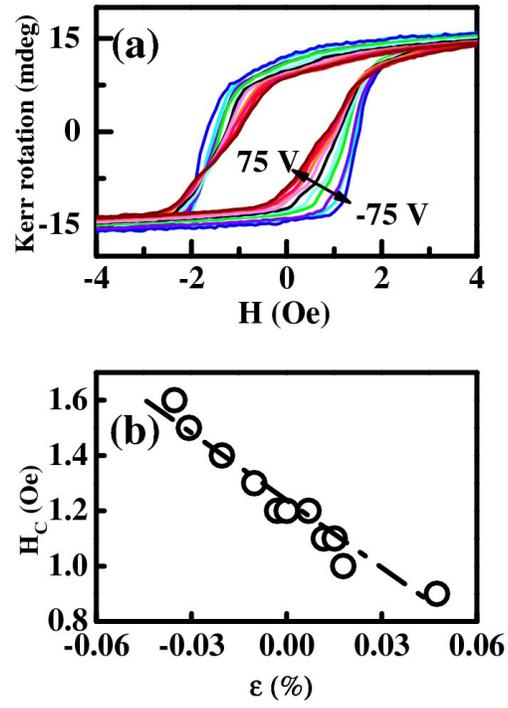

*Figure 4 Y.Y. Li et al.*

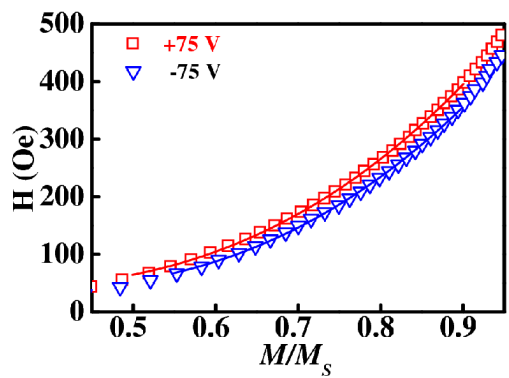

*Figure 5 Y.Y. Li et al.*